\documentclass{llncs}
\usepackage{graphics}
\usepackage{graphicx}
\graphicspath{{figures/}}
\usepackage{url}
\usepackage{float}
\floatstyle{ruled}
\newfloat{program}{thp}{lop}
\floatname{program}{\small{}Prog.}

\newcommand{\mj}[1]{}
\newlength{\figurewidth}
\begin{document}



\title{SONoMA: A Service Oriented Network Measurement Architecture}
\subtitle{TR-ELTE-CNL-2010/1, date: 17 May 2010}
\author{B\'ela Hull\'ar \and S\'andor Laki \and J\'ozsef St\'eger \and
Istv\'an Csabai \and G\'abor Vattay}
\institute{Departent of Physics of Complex Systems \\
E\"otv\"os Lor\'and University, Budapest, Hungary \\
\email{\{hullar, laki, steger, csabai, vattay\}@complex.elte.hu}}
\date{17 May 2010}


\maketitle
\begin{abstract}
To characterize the structure, dynamics and operational state of the Internet it requires distributed measurements. Although in the last decades several systems capable to do this have been created, the easy access of these infrastructures and orchestration of complex measurements is not solved. We propose a system architecture that combines the flexibility of mature network measurement infrastructures such as PlanetLab or ETOMIC with the general accessibility and popularity of public services like Web based bandwidth measurement or traceroute servers. To realize these requirements we developed a multi-layer architecture based on Web Services and the basic principles of SOA, which is a very popular paradigm in distributed business application development.
Our approach opens the door to perform complex network measurements, handles heterogeneous measurement devices, automatically stores the results in a public database and protects against malicious users as well. To demonstrate our concept we developed a public prototype system, called SONoMA.
\end{abstract}


\section{Introduction} 
\label{sec_intro}
Since the last 50 years Internet has grown from an academic experiment with several small attached networks to a highly interconnected heterogeneous system that spans several continents.  Recently it is a network of networks that consists of millions of private and public, academic, business, and government networks of local to global scope.
Besides the Internet's expansion the growing number of users and applications generate huge and more complex network traffic to be handled which poses many challenges for network operators and the network itself. As a consequence traffic control, forecasting, performance analysis and monitoring are becoming fundamental issues for network operators and interesting targets for researchers as well. 

To determine the key performance metrics needed to analyze network behavior and network traffic,
numerous independent network measurement infrastructures and testbeds have been developed and deployed all over the world. These infrastructures aim at helping researchers to examine many interesting aspects of the Internet like network topology, traffic behavior, one-way and queuing delay fluctuations or routing policies. 
Nevertheless, the way to use them is very different and sometimes not too comfortable. In general, it splits into the following key steps: first we write a measurement script then we upload it to the measurement nodes (via Web or a direct terminal connection). After that we execute it and finally we collect the results and optionally store them in a database.

We have to mention that some ISPs provide publicly accessible measurement services, which are very popular among Internet users thanks to their convenient Web based accessibility. 
Their network measurement capabilities, however, are very limited and insufficient for the research community.
One of the open questions in network research is how the flexibility of the major network measurement infrastructures can be combined with the general accessibility and popularity of these lightweight Looking Glass services.

This paper outlines a Web Service based approach for building an integrated architecture for network measurements  that is scalable, adaptable and open for scientists and other network developers, while its functionalities can be easily accessed through a standardized interface. The Service Oriented Architecture (SOA) is a very popular principle in system design and integration concerning business applications. Naturally the key components of this principle can be used in the design of network measurement architectures as well. In this paper we introduce the main concept of a SOA-like Service Oriented Network Measurement Architecture and its working prototype called SONoMA \cite{SONoMA}.

SONoMA is a common and extensible network measurement framework which proposes an alternative to define and perform distributed network experiments. 
This SOA based approach aims to decrease the required time and efforts of network experiment implementation significantly.
The two key components for achieving this goal are the conduction of complex measurements by the system and the easy invocation of the services by the web service technology.
To perform a measurement using this system there is only one thing to do: prescribe what you want to measure and then the framework will ensure to deliver the required measurement data.
Nevertheless, the above request will be disassembled to individual executable tasks in the background. Each task will be performed on a proper set of measurement nodes in a completely distributed manner.
Whilst the results are forwarded back to the user they are automatically stored in our public database, which is called the Network Measurement Virtual Observatory (VO) \cite{vo2}.
Researchers do not have to waist time developing scripts which check the status of the measurement nodes, spread the probing tasks among the nodes then collect the results and finally post process the data.

The rest of the paper is organized as follows: in Section~\ref{sec_stateart} we overview the state of the art including the prior network measurement facilities and testbeds. Section~\ref{sec_ws} briefly introduces the Web Services technology. The key concepts of our service oriented network measurement architecture are presented in Section~\ref{sec_measurements}, while its prototype implementation is detailed in Section~\ref{sec_protsys}. Section \ref{sec_case} focuses on the case studies showing how simple it is to use SONoMA to perform distributed network measurements. The final section summarizes our results.

\section{State of the Art}
\label{sec_stateart}
The idea of building global network measurement infrastructures is not new.
In the last decades numerous facilities have been developed and deployed all over the world.
The mature ones like PlanetLab \cite{planetlab} or ETOMIC \cite{etomicalapcikk} provide almost full control over their geographically dispersed measuring nodes. 
Besides network measurements they open the door to try out new network protocols and applications as well. 
This kind of freedom makes them general testbeds.
Nevertheless, this freedom is not necessary for most of the use cases required by the network measurement community.
On the other hand, it makes the development and deployment of network experiments needlessly complicated. 
and the usage of this system requires registration and other restrictions.

There are other projects like DIMES \cite{dimes}, which takes a different approach than building and maintaining a costly permanent infrastructure.
The members of this community are volunteers who choose to install the DIMES agent on their PC and these agents use the idle time of the their computers to download and perform measurement tasks.
The capabilities of these software agents, however, are very limited and the use of this system also requires registration and learning a new interface.

Scriptroute \cite{scriptroute}, in contrast with DIMES, provides a general and flexible software platform for defining network experiments easily. 
It proposes a Ruby based scripting environment which makes an ordinary PC be able to carry out network measurements instrumented remotely and safely. 
The weakness of this approach is its lacking a uniform mechanism for complex measurements which require the cooperation of a set of the measurement nodes (e.g. chirp measurements, network tomography, etc.).
Thus in this scenario users themselves have to build up their distributed measurements by synchronizing the active probing nodes.
Scriptroute is currently deployed and accessible on PlanetLab slices and its users have to implement and build up their distributed measurements in this determinate language.

Besides the above solutions there are numerous ISPs providing lightweight measurement services like traceroute or ping, which are very popular among Internet users thanks to their public accessibility. 
These services are advertised via their own Web based user interface.
The available widely heterogeneous interfaces do not facilitate to perform distributed measurements in a unified access manner. 
Furthermore, the measurement types they offer are mostly limited to ping and traceroute.


Several national R\&E networks including G\'EANT2 and Internet2 joined their forces to specify and implement a service oriented monitor infrastructure for collecting and publishing network performance data, called perfSONAR \cite{perfsonar}. 
perfSONAR is based on Web Services as well as SONoMA and they have much in common in their architecture. However there are several conceptual differences between them. 
Network operators designed perfSONAR for monitoring purposes, while our approach focuses on the problems of complex network measurements, especially active probing. 
\mj{meg valami differencia?}

SONoMA introduces a sophisticated resource allocation strategy, which differs from the ones used in existing infrastructures.
For example, PlanetLab accomplishes a time sharing approach by running several virtual machines in parallel, while ETOMIC operates a time allocation system to ensure precise timing during the experiments. 
It is obvious that most of the network monitoring probes do not need dedicated resources, whereas in certain special cases precision will still require high availability of the peripherals and resources.
Our architecture supports both resource managing mechanisms. 

As we mentioned before numerous lightweight network diagnostic services like bandwidth measuring tools are publicly available.
In contrast to the major network measurement infrastructures these basic services do not require user authorization, resource management and resource allocation.
However, their limited tool set is more than enough for most of the Internet users which makes these services very popular.
The approach discussed in this paper makes an attempt to integrate the complexity of major network experimentation facilities and the popularity of the above lightweight services. 
It provides public access with reasonable limitations for performing basic measurements and gives full access to registered users for assembling complex Internet experiments.

In addition to the above aspects we distinguish two different invocation modes in SONoMA: synchronous services for fast and immediate measurements and asynchronous ones for long-running probes.

\mj{1. guest/regisztracio}
\mj{2. ide kell?}
\mj{sync / async?}
\mj{In SONoMA two levels of measurement tasks are defined. There are atomic measurement services provided by the measurement nodes themselves and complex experiments, which are assembled from atomic ones aided by complex resource management (e.g. sending chirps between two measurement points).
Besides the above differences, the approach introduced in this paper enables both synchronous and asynchronous measurement invocation types for short and long term experiments as well.}


\section{Web Service}
\label{sec_ws}

Web Service is a technology that provides interoperability between different computers, platforms and applications to build up a distributed system. 
%
%
%
This technology has many advantages.
First, Web Services are independent form any operating systems and programming languages. Second, they offer a simplified access to remote procedures. This platform independent approach is very useful in a distributed environment, where a lot of systems interact. 
In the last decades it has became a mature technology which contains numerous standards and extensions like WS Security for using encryption and signatures to secure message exchanges.
In addition, most of the programming languages support the creation and the use of Web Services and offer tools for the automated code generation.

The Service-Oriented Architecture (SOA) is an effective distributed system development paradigm based on Web Services. In this approach the system is made up of loosely coupled services which communicate each other through the network. The deployed services can be used by other organizations and companies according to the well defined interfaces. This system design with the standardization enables the effective and strong cooperation between the different participants of a field. For example, the travel industry now has a well-defined set of services which allows software engineers to create travel agency applications easily. These products use the off-the-shelf services of different companies like other travel agencies, hotels or airline companies and improves the efficiency thanks to a big degree of automation.
\mj{jol bevezetett?}

This SOA based system design is currently under adaptation in the field of network management\footnote{e.g. ETSI TS 188 001 NGN, ITU-T M.3060}. 
Although in the literature there exists a few proposals based on the Web Services technology in connection with network measurements, a mature, well standardized solution is still missing.

\section{Architecture Overview}
\label{sec_measurements}
The idea of exploiting the advantages and flexibility of Web Service technologies in the context of network measurements is not a new one \cite{perfsonar,pam_com,icon}. 
Building and maintaining network measurement repositories and sharing experimental data \cite{vo} is also an apparent trend in both engineering and scientific community.

Our system concept tries to bind the two together.
Namely, we propose a three-tier system architecture that provides simple means to carry out a broad range of network measurements on the one hand, and on the other the system is responsible for saving all raw network measurement data in a public repository also called Network Measurement Virtual Observatory.
The idea behind collecting raw data is apparent.
It provides a unified methodology to represent and publish measurement data and it also makes reproducibility possible, for example if different new methods become available to estimate some characteristics of the network they are easily fed by historical data for the sake of comparison.

\begin{figure}
\centering\includegraphics[width=1.6\figurewidth, angle=0]{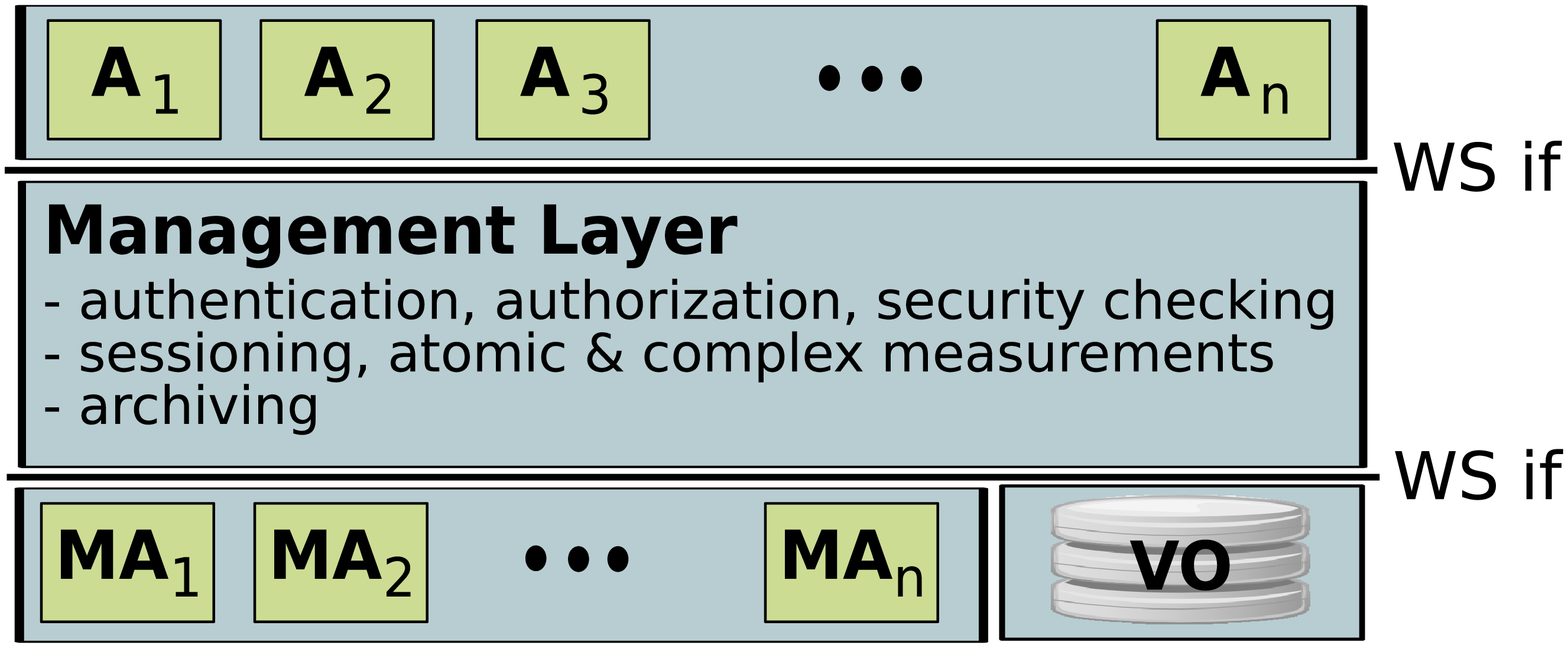}
\caption{The layout of the software architecture. Actors (A), the Management Layer and Measurement Agents (MA) form a 3-tier framework. Data are archived in the public Network Measurement Virtual Observatory (VO).}
\label{fig_tiers}
\end{figure}

Fig. \ref{fig_tiers}. shows the architecture of the system.
\textbf{Measurement Actors} (A) are the topmost entities in this model.
In the most natural case Actors are user applications implemented in any programming or scripting languages capable of integrating Web Service functionalities. 
The main advantage of this approach is that user can simply embed network measurements into their applications seamlessly. They can process their results within that framework and/or trigger signals based on the evaluation of the raw measurement data received in turn of calling the services.
At the same time, this model framework also offers a more convenient solution for the price of a less flexible way to interact with the system. 
Namely, Actors could also be nicely elaborated graphical user interfaces served via the world wide web.
In both cases a well defined programming interface is provided the Actor to contact the middleware, the Management Layer.

\begin{figure}
\centerline{
\includegraphics[width=1.5\figurewidth, angle=0]{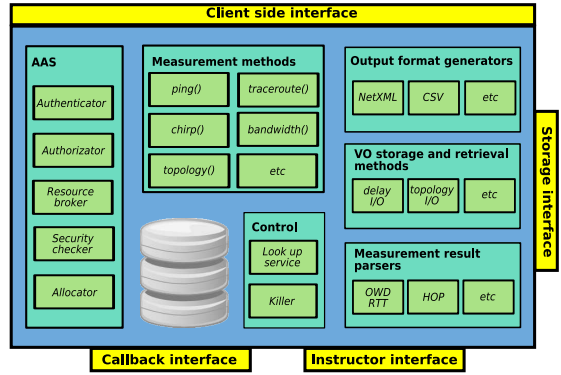}
}
\caption{The internal structure of the Management Layer middleware.}
\label{fig_modules}
\end{figure}

The \textbf{Management Layer} (ML) is the second layer in this model and is responsible for implementing and serving the following fundamental operations: i) accounting both Actors and Measurement Agents, ii) authenticating Actors, authorizing requests and checking them against misuse, iii) handling measurement sessions, scheduling and composing experiments.
These tasks are detailed in Fig. \ref{fig_modules}.
\mj{On one hand through this middle ware Actors can invoke complex measurements, this layer assign each sub-task to the measurement units then collects the date and provides the result in a proper format and store it in a database. On the other hand it controls the access, helps to prevent attacks or malicious use of the resources.}

The \textbf{Measurement Agents} (MA) and the \textbf{Virtual Observatory} (VO) reside at the lowest level of the proposed model.
Conceptually, any network entities implemented the required web service interface can be a MA.
However, it does not need to implement all the available network measurements, the ML keeps track of all the MAs and their abilities.

\mj{lehet, hogy ez mar a megvalositashoz kellene?}
In the proposed system we apply two orthogonal classifications of the network measurements.
From user's point of view we differentiate between synchronous and asynchronous measurement calls.
In the former case network measurements are parametrized in such a way, that results are available within a few seconds.
Thus the measurement results will be returned to the Actors directly and the code calling the service functions will be blocked until the measurement ends. 
In the latter case experiments are parametrized for long runs and only a measurement reference is returned, which makes it possible to try and retrieve results later, when they are available. 
In both cases the raw data produced by MAs are simply stored in the VO.

The other classification is based on the measurement complexity.
We define atomic measurements and complex measurements.
Atomic measurements are the simplest building blocks of any network measurement (e.g. traceroute, ping, etc.). 
The collection of atomic measurements (or a part of all the possible ones) are offered by MAs.
However, complex measurements are realized at ML level.
In most cases they combine several basic measurements running on one or more MAs (e.g, bandwith measurement, network tomography, geographical localization, etc.).
Then ML has to collect and match pieces of information properly together, just like aligning a probe packets time stamps upon emission and reception.
In addition a complex measurement may contain post processing, data formatting and evaluation phases as well.
In this case the aggregated or evaluated values like queuing delay histograms or location estimates will be stored in the VO as well as the raw measurement data like one-way delays or round-trip times required for the calculation of these complex network characteristics.


The model described here enables both time-sharing and time-reserving measurement approaches.
It is dependent on the requirements of the given network measurement, which are formulated in measurement rules by the ML.
For example, topology discovery and echo measurements do not require dedicated resources, thus the implementations of \texttt{traceroute} and \texttt{ping} run in time-sharing mode, whereas network tomography measurement \cite{tomo}, which requires the generation and sending of highly correlated IP packets and precise time stamping is definitely executed in time-reservation mode.

\section{A Prototype System}
\label{sec_protsys}
Testing our concept and showing its advantages we have implemented the key elements of the proposed architecture including the Management Layer (interfacing to VO), two types of Measurement Agents, and a few Actor examples, as well.
We separated the different functionalities and worked out a clear design, which makes the system be able to adopt to the new demands plainly.
This section overviews the building blocks of the realized components of the system and the thoughts behind their implementations.

\subsection{Measurement Actors}
Users are free to realize their Actors in the programming environment they use everyday, since web service technology is supported by almost all programming languages.
Actors are offered the service descriptor file (WSDL in document literal format) from which the interface code can be generated automatically.
Including the interface code in the source of the Actor all measurement services are available through simple function calls (see \ref{sec_case} for examples).

For demonstration purposes we developed a simple web accessible application where a few synchronous measurements are available that require no extra analysis of the data.\cite{SONoMA}
Here the user has freedom to modify some of the parameters of the measurements via input boxes and pull down menus.

Naturally a simple web form is incapable of taking the full power of delivering asynchronous long measurements, so users in need for them will need to fall back to the use of the WSDL.
In Table \ref{func_list} we enlist the currently implemented measurement services and also indicate those that we plan to extend with in the near future.

\begin{table}
\caption{The list of the supported measurement services (* marks services that are currently under implementation)}
\begin{center}
\begin{small}
\begin{tabular}{l|p{8cm}}
\hline
Function name & Short description\\
\hline \hline
getVersion & queries the current version of the ML \\
requestSession & opens a new session\\
closeSession & closes the given session\\
getNodeList & returns the list of measurement nodes according to the given type parameter\\
shortPing & synchronous ping measurement \\
longPing & asynchronous ping measurement\\
paralellPing & asynchronous ping measurement towards different destinations\\
ensamblePing & asynchronous ping measurement towards different destinations from different sources\\
shortTraceroute & synchronous treaceroute measurement\\
longTraceroute & asynchronous traceroute measurement\\
paralellTraceroute & asynchronous traceroute measurement towards different destinations\\
ensambleTraceroute & asynchronous traceroute measurement towards different destinations from different sources\\
shortChirp & synchronous chirp measurement from a source MA to a destination MA \\
longChirp & asynchronous chirp measurement from a source MA to a destination MA \\
getAvailableBandwidth & performs bandwidth measurement between two measurement nodes\\
shortTrain & synchronous back-to-back packet train sender from a source MA to several destination MAs \\
longTrain & asynchronous back-to-back packet train sender from a source MA to several destination MAs \\   
getResults & returns the results of a terminated asynchronous measurement \\
getProcessInfo & queries the status of a submitted asynchronous measurement \\
topology\footnotemark[1] & performs traceroute measurements between a set of MAs and gives back the topology graph \\
queuingDelayTomography\footnotemark[1] & computes and gives back the distribution of queuing delay fluctuations on the topology spanned by the measurement nodes \\
queuingDelayVariance\footnotemark[1] & computes the variance of queuing delay fluctuations on the topology spanned by  the measurement nodes \\
geographicalLocalization\footnotemark[1] & performs delay and topology measurements to localize the given IP address \\
pcapSender\footnotemark[1] & sends out a general packet pattern described by a standard pcap file \\
\hline
\end{tabular}
\end{small}
\end{center}
\end{table}\label{func_list}

\subsection{Management Layer}
An elaborate view of our Management Layer implementation is depicted in Fig. \ref{fig_modules}.
The ML provides four operational interfaces, all dedicated for different purposes.
i) The \textit{Client side interface} is an input/output interface offered for the Actors. 
All the functionalities related to requesting measurement sessions, carrying out the measurements themselves and methods of data retrieval are described here in a WSDL description. 
The rest of the interfaces are hidden from the Actors.
ii) The \textit{Instructor interface} as for its functionality is similar to the Client side interface residing at a level lower, between the ML and the MA.
Note that the WSDL description of this interface is constrained to the atomic measurement methods only.
iii) To avoid errors originating from a bad experiment duration guess the \textit{Callback interface} is provided to MAs, where long processes can notify the middleware of process termination and trigger the Management Layer to take the necessary steps of data retrieval.
iv) The raw data of each and every atomic measurement is loaded to the VO via the \textit{Storage interface}.

Considering the time line of a measurement, the modules of Management Layer will belong to three sets: i) services and tasks to be invoked before measurement, ii) services and modules, which are responsible for managing measurements, and iii) modules that pre-process and store raw measurement data.

The first group, the \textit{Authentication, Authorization \& Session handling} (AAS) collect the modules to call prior to measurements.
The tasks of \textit{Authenticator} and \textit{Authorizator} is self-explanatory.
Authenticated Actors are granted a session and a privilege schema.
The session binds all the network measurements of an Actor together, which is also represented in the databases of the VO.
The session also encapsulates the overall requests of an Actor such as the format it expects to receive the measurement data.
The privilege schema, which is checked during measurement authorization, distinguishes between different Actors, e.g. an Actor authenticated as guest has no privileges to run asynchronous long measurements and quotas are introduced on the frequency they use the system, whereas respected Actors have larger freedom to exploit the capabilities of the system.
The \textit{Allocator} and the \textit{Resource broker} modules are responsible for checking if the chosen MA is capable of a given experiment and whether the required resources are available. If the experiment requires a time-reserving operation then a certain estimated time interval will be allocated and parallel measurement requests will be omitted.
The \textit{Security checker} uses heuristics to filter out the unlucky combination of the experiment parameters, which may lead to malicious or blocking traffic.
It also maintains a gray list of MAs that have some constraints on their atomic measurement methods, e.g. operators of MAs connected to low speed links typically would not like probe packet generation at a high rate.

The second set of modules, \textit{Control} and \textit{Measurement methods} manage the network measurements.
Lookup service provides Actors an up-to-date status information of the set of MAs against different filtering rules, like the nodes that are available and/or are capable of a certain measurement type or may inform Actor when an MA is free again in case it is busy at present.
It is also the Lookup service's duty to handle signals from the Callback interface and to initiate data retrieval from the given nodes.
The Killer service can be used to hang up a measurement and free allocated resources just in case the Actor is not interested in the result any more.
The Measurement methods interface contains low level measurements (like shortPing()) offered by MAs and complex ones implemented at Management level (like getAvailableBandwidth()).

The third set of modules operates on the raw measurement data.
\textit{Measurement result parsers} provide classes for each basic measurement type to read, represent and analyze their responses.
\textit{VO storage and retrieval methods} build up database connection on demand and embody queues to store data.
In addition, data retrieval after a long measurement is manifested here.
\textit{Output format generators} serialize the raw measurement data in the requested format (e.g. NetXML, CSV, etc.).


\subsection{Measurement Agent}
In the current version of the prototype system Measurement Agents are running on PlanetLab slices  and on the Active Probing Equipments (APE) installed at OneLab \cite{ape} sites. 
While the agents running on PlanetLab are based on ordinary PC architecture, APE is an embedded, real-time system which provide an active probing platform with GPS synchronized precise packet time stamping hardware add-on.
Implementing the two kind of agents required different programing libraries and environments, but ML and the system users do not experience any differences since the web services technology hides the details.


\section{Case Studies}
\label{sec_case}

To demonstrate the benefits of our approach, we show two complex use cases: a bandwidth measurement and a topology discovery experiment.
In both cases the Actors codes are implemented in python.
The main steps of the program flow are i) instantiate the web service and request a session, ii) run the desired measurement and collect data, iii) close session and post-process the results.

\begin{figure}
\includegraphics[width=.9\figurewidth, angle=0]{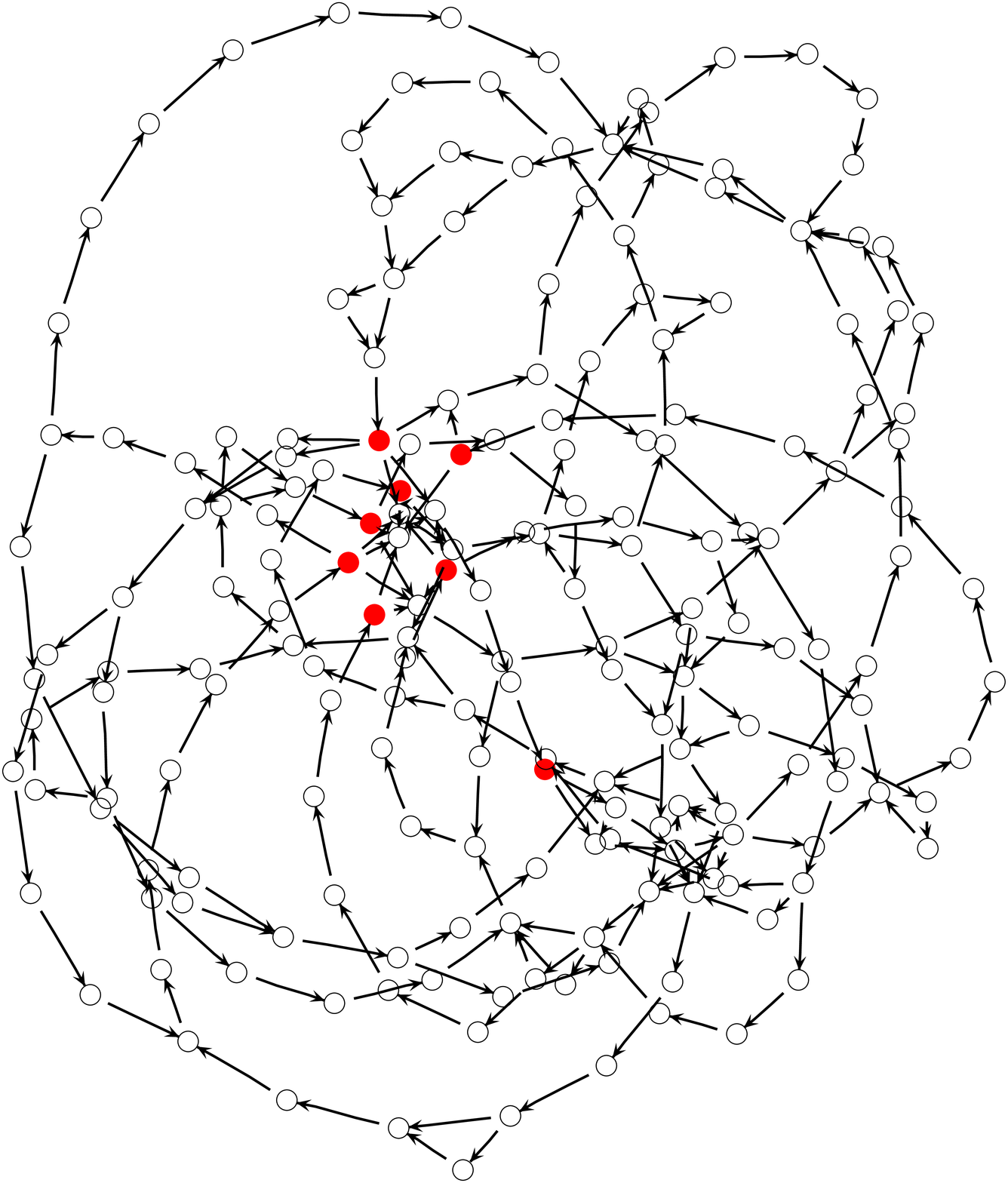}%
\\[-14em]\hspace*{1.1\figurewidth}\begin{tabular}{l|r}
\hline
Metric & measure\\
\hline \hline
Collected routes & 56\\
Mean/STDEV of route lengths & 15.4 / 4.9\\
Number of edges & 254\\
Number of nodes & 194\\
Alternative routes & 38\\
Number of load balancers & 42\\
Number of links with delay statistics & 122\\
Average delay & 20.4 ms\\
Links of $<$1 ms delay & 180\\
\hline
\end{tabular}\\[1em]
\caption{We show the output of Prog. \ref{prog_topo} and the basic statistical properties.}
\label{fig_graph}
\end{figure}

\begin{figure}
\centerline{
\includegraphics[width=1\textwidth, angle=0]{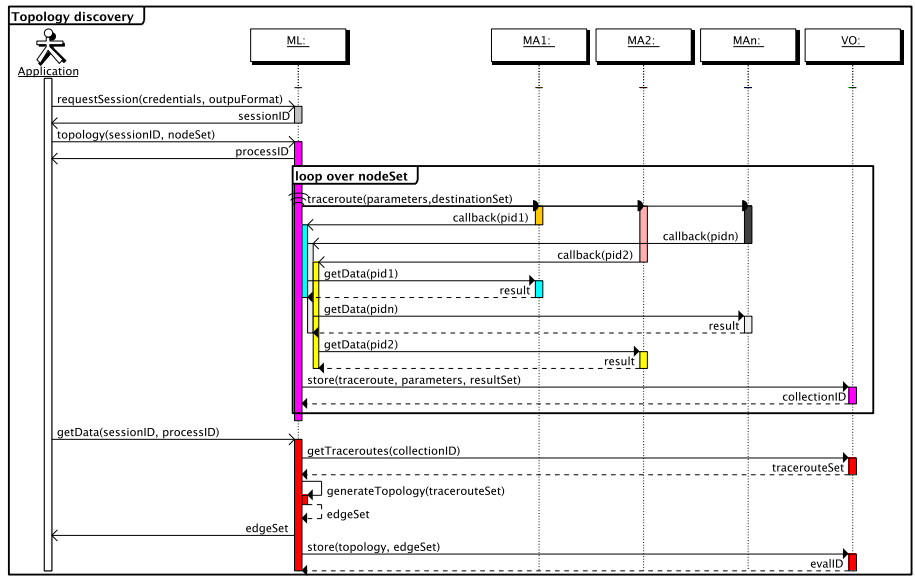}%
}
\caption{
This sequence diagram illustrates how a topology discovery is disassembled to atomic traceroute measurements running on MAs. It can be seen that once the measurements finish, the results are collected and evaluated to generate a topology map, and both raw and post processed data are stored in the VO automatically.
}
\label{top_seq}
\end{figure}

The first case (Prog \ref{prog_topo}) implements an asynchronous measurement, where we try to draw the connection graph among a set of MAs.
We note here that while requesting a session from Management Layer, user passes a description of the desired result format, i.e. comma separated values, zipped output in this case.
Next, the web method's parameters are set: the measurement nodes between which we intend to discover the network topology.
After calling the properly parametrized topology() web method an unique identifier, a processID is returned.
As it can be seen on Fig. \ref{top_seq} this complex measurement is disassembled into the execution of individual traceroute probes, whose results are pushed individually to the VO after completion.
When all partial measurements are over, the Actor is able to retrieve the set of links using the getData() method.
In calling gatData() based on the sessionID, prosessID pair, ML retrieves traceroute data from the VO repository and realizes the construction of the link set to be returned.
Using 8 nodes in the experiment we may draw the graph like in Fig. \ref{fig_graph}.

In the second case we request for a bandwidth estimate between two MAs.
Compared to the previous example, the main difference here is that we use the getAvailableBandwidth() method which represents a synchronous measurement. 
Therefore, after calling the above procedure the measurement results will be returned straight away.
However, in the background ML breaks down this complex query into two atomic measurement instances: a chirp sender and a capture process.
Next, it checks the availability of the nodes involved in the measurement, reserves them, runs the processes and collects data.
Right after the collection of the sending and receiving timestamps, ML matches them properly and stores the raw data in the VO.
In the meantime an estimation process is launched to digest raw time series and the evaluated result is returned to the user.

We hope that the introduced two examples show how easily SONoMA can be used to perform complex and distributed network measurements like topology discovery or bandwidth estimation.

\mj{
\begin{program}
{\scriptsize
\begin{verbatim}
# instantiate web service
ws = ServiceServerSOAP( "http://157.181.172.123:22222" )

# request a session
sessionID = ws.requestSession( user="guest", zipResults=False, formatResults="CSV" )

# do the measurement
bandwidth = ws.getAvailableBandwidth( sessionID, src="157.181.175.247", dst="132.65.240.38" )

# post-process data and close session
print bandwidth
ws.closeSession( sessionID )
\end{verbatim}
}
\caption{\small{}The pseudo code to call getAvailableBandwidth() web method.\label{prog_bw}}
\end{program}
}

\mj{1. lesz full mesh?}

\begin{program}
{\scriptsize
\begin{verbatim}
# instantiate web service
ws = ServiceServerSOAP( "http://157.181.172.123:8888" )

# request a session
sessionID = ws.requestSession( user="User", zipResults=True, formatResults="CSV" )

# do the measurement
(processID, expectedDuration) = \
    ws.topology( sessionID, nodeList = [ "157.181.175.247", "132.65.240.38", ... ] )
                 ...
# wait for some time & retrieve data from the VO
result = ws.getData( sessionID, processID )

# post-process data and terminate session
drawGraph( decompress(result) )
ws.closeSession( sessionID )
\end{verbatim}
}
\caption{\small{}The pseudo code of a topology discovery measurement among 8 MAs.\label{prog_topo}}
\end{program}

\section{Conclusion}
\label{sec_conclusion}
In this paper we introduced a novel network measurement architecture, SONoMA based on Web Services.
In the proposed three-tier system users reach the Measurement Agents through an intermediate Management Layer.
This middleware controls the access, offers atomic and complex measurements, hides the heterogeneity of the agents and stores the results in a public repository. This concept enables the uniform handling of active and passive measurements and allows the time sharing and the time reservation resource allocation schemata, too.
Furthermore we have presented our publicly available prototype, whose agents are currently implemented and deployed on PlanetLab nodes and on the APE platform. Finally, we demonstrated the system's benefits with some use cases.
The presented infrastructure provides ordinary and complex network measurements in easily and publicly accessible way.

Further research work needs to be done which aims at increasing the number of measurement types both at complex and atomic levels, to integrate WS-Security into the authentication and user identification process as well as to examine the performance and the efficiency of the system implementation.
In addition, the number of Measurement Agents will be extended with the precise active measurement infrastructures of the OneLab2 project and, of course, we count on other institutions to join the SONoMA platform \cite{SONoMA} as users or as Measurement Agent operators as well.
We also have to think of the adaptation of volunteer, desktop grid like agents, that worked well in other projects like DIMES.

\section{Acknowledgment}
The authors thank the partial support of the National Office for Research and Technology (NAP 2005/ KCKHA005), 
the EU ICT MOMENT Collaborative Project (G.A.No. 215225) and the EU ICT OneLab2 Integrated Project (G.A.No. 224263).

\bibliographystyle{plain}
\bibliography{tr-elte-cnl-2010-sonoma}

\end{document}